\documentclass{elsart}
\usepackage{amssymb}
\usepackage{graphicx} 
\usepackage{dcolumn}  
\usepackage{xspace}
\newcommand{\flow}{\ensuremath{\mathcal{F}_{\mathrm{FLOW}}\xspace}}

\newcommand{\gev}{\ensuremath{\mathrm{GeV}\xspace}}

\newcommand{\ra}{\rightarrow}

\begin{document}
\begin{frontmatter}
\vspace*{-3\baselineskip}
\begin{flushleft}
 \resizebox{!}{28mm}{\includegraphics{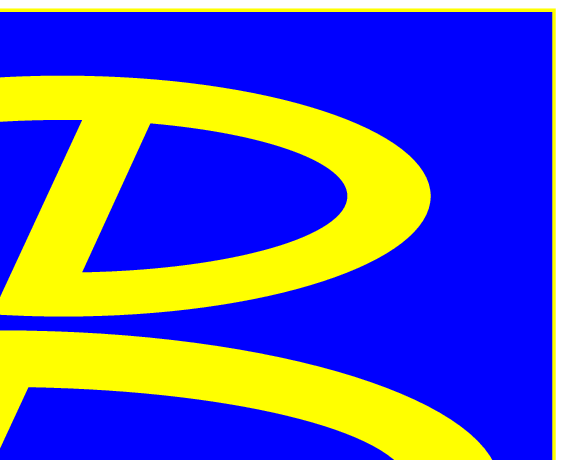}}
\end{flushleft}
\vspace*{-28mm}
\begin{flushright}
  Belle Preprint 2005-18 \\
  KEK Preprint 2005-15
\end{flushright}

\vspace*{12mm}

  \title{ \quad\\[0.5cm]
    Measurement of $|V_{ub}|$ near the endpoint of the
    electron momentum spectrum from semileptonic $B$-meson decays
  }

\collab{Belle Collaboration}
  \author[KEK]{A.~Limosani}, 
  \author[KEK]{K.~Abe}, 
  \author[TohokuGakuin]{K.~Abe}, 
  \author[KEK]{I.~Adachi}, 
  \author[Tokyo]{H.~Aihara}, 
  \author[Tsukuba]{Y.~Asano}, 
  \author[ITEP]{T.~Aushev}, 
  \author[Cincinnati]{S.~Bahinipati}, 
  \author[Sydney]{A.~M.~Bakich}, 
  \author[Melbourne]{E.~Barberio}, 
% ...

% ...
\address[Cincinnati]{University of Cincinnati, Cincinnati, OH, USA}
% ...
\address[KEK]{High Energy Accelerator Research Organization (KEK), Tsukuba, Japan}
% ...
\address[ITEP]{Institute for Theoretical and Experimental Physics, Moscow, Russia}
% ...
\address[Melbourne]{University of Melbourne, Victoria, Australia}
% ...
\address[Sydney]{University of Sydney, Sydney, NSW, Australia}
% ...
\address[TohokuGakuin]{Tohoku Gakuin University, Tagajo, Japan}
% ...
\address[Tokyo]{Department of Physics, University of Tokyo, Tokyo, Japan}
% ...
\address[Tsukuba]{University of Tsukuba, Tsukuba, Japan}
% ...
\thanks[NovaGorica]{on leave from Nova Gorica Polytechnic, Nova Gorica, Slovenia}

\begin{abstract}
  We report measurements of partial branching fractions of inclusive
  charmless semileptonic $B$-meson decays at the endpoint of the
  electron momentum spectrum. The measurements are made
  in six overlapping momentum intervals that have lower bounds
  ranging from $1.9$~GeV/$c$ to $2.4$~GeV/$c$ and a common upper bound
  of $2.6$~GeV/$c$, as measured in the centre of mass frame. 
  The results are based on a sample of 29
  million $B\overline{B}$ pairs, accumulated by the Belle detector at the
  KEKB asymmetric $e^+e^-$ collider operating on the $\Upsilon(4S)$
  resonance.
  In the momentum interval ranging from  $1.9$~GeV/$c$ to
  $2.6$~GeV/$c$ we measure the partial branching fraction
  $\Delta \mathcal{B}(B\rightarrow X_u e \nu_e)=(8.47 \pm 0.37 \pm 1.53)\times
  10^{-4}$, where the first error is statistical and the second is
  systematic. A prediction of the partial rate $R=(21.69 \pm
  3.62^{\,+\,2.18}_{\,-\,1.98})\,|V_{ub}|^2 \mathrm{ps}^{-1}$  in this momentum
  interval based on theory is calculated with input HQET parameters
  that have been derived from Belle's measurement of the $B\rightarrow X_s\gamma$
  photon energy spectrum, where the first error is
  due to the uncertainty on HQET parameters and the second error is from
  theory. Using both $\Delta \mathcal{B}(B\rightarrow X_u e \nu_e)$ and $R$ we
  find $|V_{ub}|=(5.08 \pm 0.47 \pm 0.42^{\,+\,0.26}_{\,-\,0.23})\times
  10^{-3}$, where the first error is from the partial branching
  fraction, and the second and third errors are from uncertainties in $R$.  
\end{abstract}

\end{frontmatter}


\section{Introduction}
\label{intro}
The magnitude of the
Cabibbo-Kobayashi-Maskawa (CKM) matrix element $|V_{ub}|$ is a
fundamental parameter of the Standard Model (SM). A knowledge of its
value is crucial to the
understanding of $CP$ violation within the SM, which is
underpinned by knowledge of the so-called Unitarity Triangle (UT).
Recent precise measurements of UT angle
$\phi_1$($\beta$)~\cite{Abe:2001xe,Aubert:2001nu}
have focussed attention on $|V_{ub}|$, because it determines the side of
the UT that is opposite $\phi_1$, it directly tests the Kobayashi-Maskawa
mechanism~\cite{KM} for $CP$ violation within the SM.

To date, inclusive measurements of $|V_{ub}|$ have been reported
by experiments operating on the $\Upsilon(4S)$ resonance, namely
CLEO~\cite{Fulton:1989pk,Bartelt:1993xh,Bornheim:2002du},
ARGUS~\cite{Albrecht:1989qv}, BaBar~\cite{Aubert:2003zw}
and Belle~\cite{Kakuno:2003fk}, 
and by LEP experiments
operating on the $Z$ resonance, namely L3~\cite{Acciarri:1998if}, ALEPH~\cite{Barate:1998vv}, 
DELPHI~\cite{Abreu:2000mx} and OPAL~\cite{Abbiendi:2001qx}.

The value of $|V_{ub}|$ can be extracted from the measured rate of  
charmless semileptonic $B$-meson decays $B \rightarrow X_u e \nu_e$ 
in a kinematic region that 
has to be chosen to minimise the impact of the large
background from the charmed semileptonic $B$-meson decays $B \rightarrow
X_c e \nu_e$. One such region is at the endpoint
of the lepton momentum spectrum: in the rest frame of the decaying $B$
meson, leptons from $B \rightarrow X_c e \nu_e$ decays attain a
maximum momentum of $2.31$ GeV/$c$ while for
$B \rightarrow X_u e \nu_e$ decays the maximum is $2.64$ GeV/$c$.

In this paper we report measurements of partial branching fractions
to inclusive charmless semileptonic $B$-meson decays
from the yield of electrons and positrons in six overlapping
momentum intervals. The intervals have lower limits
commencing at $1.9\,\mathrm{GeV}/c$ through 
$2.4\,\mathrm{GeV}/c$ incremented in steps of
$0.1\,\mathrm{GeV}/c$ and a common upper limit
of $2.6\,\mathrm{GeV}/c$, as measured in the centre of mass
(CM)\footnote{The rest frame of the $\Upsilon(4S)$ is equivalent to the centre of
  mass frame.} frame.

We use two methods to extract $|V_{ub}|$, one that has been
standard practice~\cite{Group:2004cx} (DFN),
and one that has been recently developed~\cite{Bosch:2004th,Neubert:2004dd,Bosch:2004cb,Neubert:2004cu,Neubert:2004sp}(BLNP). 
The DFN method involves extrapolation from a partial
to a full branching fraction using
the DeFazio-Neubert prescription             
with given shape function parameters~\cite{DeFazio:1999sv}
followed by a routine             
to translate the full branching fraction
to $|V_{ub}|$~\cite{Group:2004cx}.                           
The BLNP method in contrast to the DFN method provides a more
systematic treatment of
shape function effects by incorporating all known contributions,
includes power corrections, uses an improved
perturbative treatment and directly relates the $B \rightarrow X_u e
\nu_e$ partial rate to $|V_{ub}|$. 
For both methods we use values for the parameters of the shape function
that were determined using the  $B\rightarrow
X_s\gamma$ photon energy spectrum measured by
Belle~\cite{Koppenburg:2004fz}\footnote{The use
  of the photon energy spectrum from inclusive radiative $B$-meson decays in
  determining the $b$-quark shape function or distribution function
  was first discussed by Bigi \emph{et}
  al~\cite{Bigi:1993ex} and Neubert~\cite{Neubert:1993um}.}.
In the case of DFN the procedure and measurements are given in
Ref.~\cite{Limosani:2004jk}. The same procedure was
updated with predicted shapes of the $B\rightarrow
X_s\gamma$ photon energy distributions from Ref~\cite{Lange:2005yw} to
yield values of shape function parameters relevant to the BLNP method.
The latter are equivalent to HQET parameters in the shape function
scheme~\cite{Neubert:2004sp}. 

\section{Detector and Data Sample}
The results reported here are based on data collected with the Belle
detector at the KEKB
asymmetric energy $e^+e^-$ collider~\cite{KEKB}.
The Belle detector is a large-solid-angle magnetic spectrometer that
consists of a three-layer silicon vertex detector (SVD),
a 50-layer central drift chamber (CDC), 
an array of aerogel threshold \v{C}erenkov counters (ACC), 
a barrel-like arrangement of time-of-flight scintillation counters (TOF), 
and an electromagnetic calorimeter comprised of CsI(Tl) crystals (ECL)
located inside a superconducting solenoid coil 
that provides a 1.5~T magnetic field.  
An iron flux-return located outside of the coil is instrumented 
to detect $K^0_L$ mesons and to identify muons (KLM).  
The detector is described in detail elsewhere~\cite{bellenim}.
We use 27.0 fb$^{-1}$  and 8.8
fb$^{-1}$ integrated luminosity samples taken at (ON) and 60 MeV below (OFF) the $\Upsilon(4S)$ resonance energy, respectively.
The ON sample consists of 29.4 million $B\overline{B}$ events. Unless explicitly
stated otherwise, all variables are calculated in the CM frame.
  
\section{Data Analysis}
The procedure for this analysis largely follows that of CLEO~\cite{Bornheim:2002du}, and
consists of examining the spectrum of
electron candidates with momentum in the range
$1.5-3.5~\mathrm{GeV}/c$, 
which includes both signal and sideband regions.
We initially chose and optimised our selection criteria 
for the momentum region,
$2.2-2.6~\mathrm{GeV}/c$.
For ease of explanation we discuss the experimental procedure for
this momentum interval and later describe the slight
differences in the signal extraction for other momentum
intervals. 

In the CM frame the kinematic endpoint for decays of the
type $B\rightarrow X_ce\nu_e$, including the non-zero $B$
momentum and detector effects, is 2.4 GeV/$c$.
The $B\rightarrow X_u e\nu_e$ signal is extracted in the momentum
region 
$2.2-2.6~\mathrm{GeV}/c$ (HI), while a lower range, from
$1.5-2.2~\mathrm{GeV}/c$ (LO),
is examined to evaluate the contribution from $B\rightarrow X_c
e\nu$, which is then extrapolated to the HI region.

The uncertainty on the fraction of $B\rightarrow X_u e \nu$ within the
HI region is a major source of systematic
error for the determination of the branching fraction and $|V_{ub}|$.
For choosing and optimising selection criteria we use
a sample of events containing $B\ra X_u e \nu_e$ decays, generated via
Monte Carlo simulation and based on a model described in
Ref.~\cite{Scora:1995ty} (ISGW2), which predicts the form factors
and branching fractions of the many exclusive charmless semileptonic
$B$-meson decay channels that form the sample.
We also generate samples based on an
inclusive $B\rightarrow X_u e \nu_e$ model,
according to the prescription of DeFazio and
Neubert~\cite{DeFazio:1999sv}, with shape function (SF) parameters
that correspond to the residual
$B$-meson mass\footnote{$\Lambda^\mathrm{SF}=M_B-m_b$; \\where $m_b$
  and $M_B$ are the masses of the $b$-quark and $B$-meson,
  respectively.} and the average momentum squared of the $b$-quark
inside the $B$-meson set to $\Lambda^\mathrm{SF}=0.659$ GeV/$c^2$ and
$-\lambda^\mathrm{SF}_1=0.400\,\gev^2/c^2$,
respectively. These values were determined from the
photon energy spectrum in $B\rightarrow X_s \gamma$
decays measured by Belle~\cite{Limosani:2004jk}.
To examine the extent to which our results may vary
due to uncertainties in $\Lambda^\mathrm{SF}$ and
$\lambda^\mathrm{SF}$, we also generate four samples
with parameters that define the long and short axes of the $\Delta\chi^2=1$
contour in the
$(\frac{\Lambda^\mathrm{SF}}{\gev/c^2},\frac{\lambda^\mathrm{SF}_1}{\gev^2/c^2})$
plane, 
corresponding to $(0.614,-0.231)$, $(0.736,-0.714)$, $(0.719,-0.462)$ and
$(0.635,-0.483)$.

\subsection{Event Selection}
To select hadronic events we require the multiplicity of
tracks to be greater than
four and the primary event vertex to be within $1.5$ cm radially and $3.5$ cm 
longitudinally from the detector origin.
We make further requirements based on quantities
calculated in the CM frame --  that
the sum of cluster energies in the ECL satisfies
$0.18\sqrt{s}<E_{\mathrm{ECL}}<0.80\sqrt{s}$
where $\sqrt{s}$ is the CM collision energy,
that the visible energy $E_{\mathrm{vis}}$ be at least $0.50\sqrt{s}$,
that the absolute sum of longitudinal track and photon momenta be
less than $0.30\sqrt{s}$, that the heavy jet mass be either at least
$0.25\times E_{\mathrm{vis}}$ or greater than $1.8$ GeV/$c^2$, and that the average
cluster energy be less than $1$ GeV.
We also require that the ratio $R_2$ of the second to the zeroth Fox-Wolfram
moment~\cite{Fox:1978vu} be less than $0.5$.

\subsection{Electron spectrum}
Charged tracks are reconstructed from hits in the SVD and the CDC. 
Tracks are required to pass within a short distance from the
interaction point (IP) of the $e^+e^-$ collision, where
the $B$-meson decays promptly.  
For improved data and MC agreement,
tracks must be within the acceptance of the barrel part of the
ECL; $-0.63 < \cos\theta_{\mathrm{lab}} < 0.85$, where $\theta_{\mathrm{lab}}$
is the polar angle measured in the laboratory frame with respect to
the direction opposite to that of the positron beam. 
Tracks are identified as electrons on the basis of a matching energy
cluster in the ECL, and, subsequently, upon the ratio of ECL-measured
energy to CDC-measured track momentum, transverse shower shape in the ECL,
ionisation energy loss in the CDC, and the ACC light yield~\cite{Hanagaki:2001fz}.
Given the track requirements, electrons with momenta in the range
$1.5-2.6~\mathrm{GeV}/c$ are positively identified
with a probability
of $(94.0\pm 1.5)$\% while pions are misidentified as electrons
with a probability of $(0.13\pm 0.01)$\%, as measured using samples of reconstructed
$J/\psi\rightarrow e^+e^-$ and $K^0_S\rightarrow \pi^+\pi^-$ decays,
respectively. 

To reduce the contribution of electrons from $J/\psi$ and $\psi(2S)$ 
decays and photon conversions, our candidate electrons are paired with
oppositely charged tracks identified as electrons
in the event and rejected if their mass falls within either  
$J/\psi$, $\psi(2S)$ or $\gamma$ mass windows,
defined as $[3.025,3.125]\,\mathrm{GeV}/c^2$,
$[3.614,3.714]\,\mathrm{GeV}/c^2$
and $[0,0.1]\,\mathrm{GeV}/c^2$, respectively. 
The photon conversion veto has the additional effect of removing electrons
from $\pi^0$ Dalitz decays.
The yields of candidates that do not pass the $J/\psi$ veto
requirement are compared in data and MC to determine a
normalisation factor for MC-estimated backgrounds, which are described
below.  

Since the dynamics of the hadronic part of $B\rightarrow X_u e\nu_e$ is
not well established, it is important that selection requirements
retain acceptance over a wide range of $q^2\equiv(p_e+p_\nu)^2$
(dilepton invariant mass squared) in order to minimise model dependence.
Additional event requirements are designed to reduce contributions
from continuum
($e^+e^-\rightarrow q\bar{q}$ where $q=u,d,s,c$)
and QED-related processes (including two
photon and tau-pair events) without introducing a $q^2$ bias.
A set of ``energy flow'' variables is formed by grouping detected
particles in bins of $0.05$  in $\cos\theta$, where $\theta$ is the
particle angle with respect to the candidate electron,
and taking the energy sum in each bin. The energy flow in
the backward direction $-1.00<\cos\theta<-0.95$ is not used,
as it is found to disproportionately reduce the acceptance in the low $q^2$ region. 
A Fisher discriminant, denoted \flow, is constructed from the remaining energy
flow variables with coefficients chosen to best separate signal from
continuum events. We also make use of a $b$-quark rare decay tag
variable, denoted $\mathcal{K}$, and calculated as
\begin{equation}
\mathcal{K} = Q(e)\left(N(K^+)-N(K^-)\right),
\end{equation}
where $Q(e)$ is the charge of the candidate electron, and $N(K^\pm)$ are
the number of tracks identified as positively and negatively charged
kaons in an event,
respectively.
$\mathcal{K}$ exploits the presence of lepton-kaon charge
correlations evident in $B\overline{B}$ events wherein one $B$-meson decays via a
$b\rightarrow u e \nu_e$ transition whilst the other $B$-meson decays typically
via $\overline{b}\rightarrow \overline{c} \rightarrow \overline{s}$
transitions,  thereby resulting
in, on average, a net strangeness or kaon charge that is
correlated to the charge of our
candidate electron. The correlation does not exist in continuum events
nor in $B\overline{B}$ events that do not involve
the $b\rightarrow ue\nu_e$ transition. 
Charged tracks are identified as kaons by utilising specific ionisation
energy loss measurements made with the CDC, light yield
readings from the ACC, and time of flight information from the TOF. 
The average kaon identification efficiency and fake rate in the
momentum range $0.5$--$4.0\,\gev/c$, as measured in the laboratory frame, are $(88.0 \pm
0.1)\%$ and ($8.5 \pm 0.1$)\%, respectively.

To preserve signal efficiency, the selection requirements on \flow\, are chosen
differently for three cases of $\mathcal{K}$: 
$\mathcal{K}>0$; $\mathcal{K}=0$; and $\mathcal{K}<0$. The cut values are chosen
to optimise the figure of merit $S/\sqrt{S+B}$, where $S$ and $B$ reflect the signal and
continuum background expectation, respectively,
as estimated from MC events, assuming
the branching fraction measured by CLEO~\cite{Bornheim:2002du}.
The $\mathcal{K}$ dependent cuts on \flow\, 
reduce continuum backgrounds by 97\% while retaining
33\% of the $B \ra X_u e\nu_e$ signal.

To further suppress QED-related continuum backgrounds,
the cosine of the angle between the thrust axis of the
event and the $e^-$ direction $\cos\theta_{thr}$, is required to be less than
$0.75$. Crucially, the thrust axis calculation includes the missing
momentum as a component. Missing momentum is calculated as the difference between the
momentum of the beams and the sum of the observed track and
cluster momenta. Placing a constraint on $\cos\theta_{thr}$ was found to bias the $q^2$
distribution in signal events less than a constraint imposed on the direction of
missing momentum, which has been
previously used by CLEO~\cite{Bornheim:2002du}.
The requirement on $\cos\theta_{thr}$ reduces QED-related
continuum backgrounds by 50\% with a signal
inefficiency of 10\%.   

\begin{figure}[h!]
\includegraphics[width=1.00\textwidth]{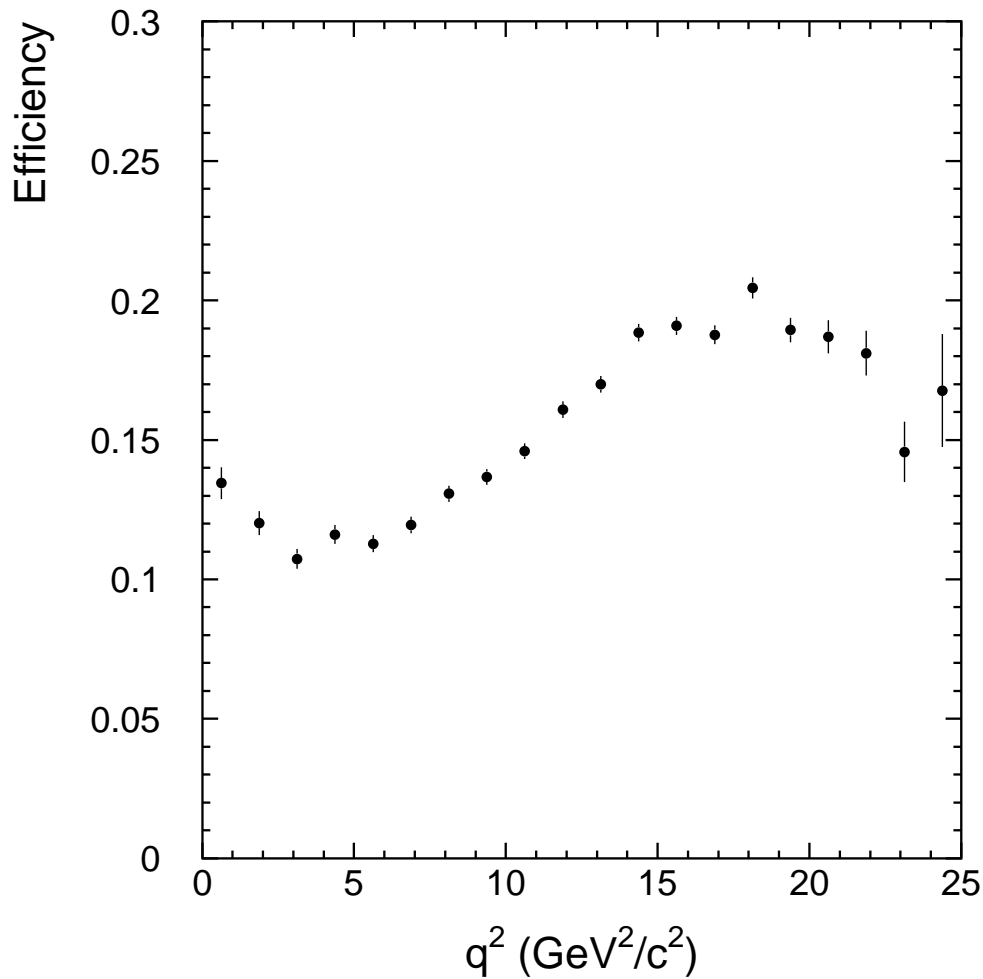}
\caption{Acceptance of the selection requirements as a function of 
generated $q^2$ for events containing electrons in the region 2.2-2.6~GeV/$c$ 
from $B\rightarrow X_u e\nu_e$ decay.}
\label{fig:2}       
\end{figure}

The acceptance of the selection requirements as a function
of generated $q^2$ for events containing electrons in the momentum interval
$2.2-2.6~\mathrm{GeV}/c$ from
$B\rightarrow X_u e\nu_e$ decay is shown in Figure~\ref{fig:2}.

\subsection{Backgrounds}
Sources of background for $B\rightarrow X_u e\nu_e$ include continuum events,
hadrons misidentified as electrons (``fakes''), decays $B\rightarrow
X_c e\nu_e$, and various secondary decays of $B$-mesons.
We describe below our evaluation of each as well as our procedures
for evaluating the associated contributions to the systematic uncertainty.

The continuum contribution is evaluated using the OFF data set.
To account for the small difference in the momentum spectra due to the beam energy
difference (0.6\%), the electron momenta in OFF data are scaled by the ratio
of ON to OFF CM energies.  
The yields in OFF data are then scaled by the factor $\alpha=3.005 \pm
0.015$, determined
by the ON to OFF ratio of Bhabha event yields in the barrel ECL.
As a check of this procedure, we measure the yields in the
momentum range
$2.8-3.5~\mathrm{GeV}/c$, 
above the kinematic maximum for $B\overline B$ events.
The resulting signal of $85 \pm 93$ electron candidates is consistent
with zero, as expected. We assign systematic uncertainties based on a
MC study of the detector response to Bhabha events (0.4\%), and the
discrepancy with $\alpha$ as calculated, similarly, with dimuon events
(0.2\%).  

The remaining  contributions to background are from $B\overline B$ events
and are estimated using a large Monte Carlo simulated sample of generic $B\overline
B$ events~\cite{BelleMC} that contains roughly three and a half
times the number of $B\overline B$ events in the ON sample.
The MC yield due to fakes from charged pions is corrected for the
difference in fake efficiency measured by data and MC samples of
$K^0_S\rightarrow \pi^+\pi^-$ decays.
The error on the correction ($\sim 30\%$) is assigned as the systematic
uncertainty on the yield from pions. Additional minor contributions from
kaons, protons and muons to the overall fakes yield are conservatively
assigned systematic uncertainties of 100\%.

Of the real electrons, those not from primary $B\ra X_c e \nu_e$,
secondary backgrounds, are
estimated using the Monte Carlo simulated generic $B\overline B$ event sample with electronic branching 
fractions of $D^0$, $D^+$, $J/\psi$, $\psi(2S)$, $D_s$ and $\tau$ assigned 
according to the current world averages~\cite{Eidelman:2004wy}.
To avoid any bias from a possible mis-modelling of data in MC
we use the vetoed $J/\psi$ sample to measure the normalisation factor
for both fake and secondary background MC yields. This factor is calculated
from a fit of the  MC-simulated momentum spectrum of vetoed
electron candidates from $J/\psi$ in $B$-meson decays to the equivalent spectrum
obtained from the data.  
This sample is statistically independent of the final event sample, and, moreover, may not
contain neutrinos from primary $B$-meson decays, which is the case
for events providing fake and secondary backgrounds.
Contributions from secondary electrons are assigned systematic
errors based on the electronic branching fraction uncertainties and 
the difference between the MC normalisation calculated as described
above and that according to the number of $\Upsilon(4S)$ events (6\%).
Overall, the latter uncertainty has a less than 0.5\% effect on the
eventual signal yield.

The spectrum from $B\ra X_c e \nu_e$ is
modelled using three components:  
$X_c = D$ (HQET~\cite{hqet}); $D^*$ (HQET~\cite{hqet}); and higher resonance charm meson 
states $D^{**}$ (ISGW2~\cite{Scora:1995ty}). 
To improve the agreement with data, we re-weight the $D$
and $D^*$ components according to $q^2$ in order to 
match spectra generated with world average values of form
factors~\cite{Eidelman:2004wy}.
The ratio of $D$ to $D^*$ branching fractions is
fixed according to their world average measurements~\cite{Eidelman:2004wy}.
The proportion of the $(D+ D^*)$ component with respect to the $D^{**}$ component is
determined from a binned maximum
likelihood fit~\cite{Barlow:dm} of the
ON data in the LO region\footnote{The sum of the yields of the components in the fit is
  constrained to equal the ON data yield.}, 
where the $B\rightarrow X_u e \nu_e$ component is 
modelled using the inclusive model described earlier and fixed such that 
$\mathcal{B}(B\rightarrow X_u e \nu_e)=(0.25\pm 0.02)\%$~\cite{Group:2004cx}.
For the $D^{**}$ sub-components, $D_1$ and $D^*_2$ we set
$\frac{\mathcal{B}(B \rightarrow D_1 e \nu_e)+\mathcal{B}(B \rightarrow D_2^* e\nu_e)}{\mathcal{B}(B\rightarrow D^{**} e \nu_e)}=0.35 \pm 0.23$, which
has been determined by averaging maximum and minimum
assessments of their respective world average branching
fractions~\cite{Eidelman:2004wy}.
Semileptonic spectra are also re-weighted
to include the effect of QED radiative corrections as calculated with
the PHOTOS algorithm~\cite{Barberio:1993qi}.
It has been observed that the KEKB collision energy has variations of 
$\mathcal{O}(1\,\mathrm{MeV})$ over time and that this results in a measurable
variation of the $B$-meson momentum over the running period of our ON
data sample.  
As our Monte Carlo generator assumes a fixed energy, we apply
a shift to the reconstructed momentum in the MC to correct for the difference.
The correction depends on the beam energy measurement in the same run
period as our ON data set, which is made using a fully
reconstructed $B$-meson decay sample. 
All spectra for backgrounds other than 
$B\rightarrow X_c e \nu_e$ are derived from the generic
$B\overline{B}$ MC sample and handled in the same manner as for the HI region.
The goodness-of-fit as estimated by the $\chi^2/\mathrm{ndf}$
gives $17.8/13$. We use this fit result to not only determine the
$B\rightarrow X_c e \nu_e$ background level in the HI
region ($2.2-2.6~\mathrm{GeV}/c$), but also
simultaneously in the signal regions defined as
$2.3-2.6~\mathrm{GeV}/c$ and
$2.4-2.6~\mathrm{GeV}/c$.

The same procedure as described above is repeated for three additional
HI regions $1.9-2.6~\mathrm{GeV}/c$,
$2.0-2.6~\mathrm{GeV}/c$, and
$2.1-2.6~\mathrm{GeV}/c$.
In each case the
LO region is adjusted such that its upper bound equals the
lower bound of the HI region, giving respective LO regions of: $1.5-1.9~\mathrm{GeV}/c$;
$1.5-2.0~\mathrm{GeV}/c$; and $1.5-2.1~\mathrm{GeV}/c$. The $\chi^2/\mathrm{ndf}$ for fits
in these LO regions are: $6.8/7$; $11.9/9$; and $13.9/11$,
respectively.

The systematic error on our measurement due to the uncertainty
in the $B\rightarrow X_c e \nu_e$ shape and relative normalisations is
estimated by varying the parameters
fixed in the fit by their individual uncertainties, as described above. The
maximum deviation observed from either an upward or downward variation 
is assigned as the systematic error. 
We also calculate uncertainties for cases of: no QED radiative
correction; an ISGW2 modelled $B \rightarrow X_u e \nu_e$ spectrum
shape~\cite{Scora:1995ty}; and the inclusion of a non-resonant
$B \rightarrow D^{(*)} \pi e \nu_e$ (Goity and
Roberts~\cite{Goity:xn}) decay component in the fit. 
CLEO included the $B \rightarrow D^{(*)} \pi e \nu_e$ component
in their standard fit~\cite{Bornheim:2002du}, but in our case,
the shape of its momentum
spectrum bears too close a resemblance to that of the $B \rightarrow
D^{**} e \nu_e$ component. If both $D^{(*)}\pi$
and $D^{**}$ components are included in the fit the $D^{**}$ component
floats to zero. 
This is
clearly contrary to observation, given the measured inclusive branching
fraction $\mathcal{B}(B \rightarrow
D^{**} e \nu)=(2.70\pm 0.70)\%$~\cite{Eidelman:2004wy}.

The systematic that has the greatest effect on the $X_c$ background
estimation in the HI 
region is the uncertainty on the $D^*$ form factor, which has been
obtained by varying the form factor slope parameter $\rho^2$ within
its uncertainty. The net
systematic uncertainty is calculated as a sum in quadrature of the
individual systematic uncertainties.
Table~\ref{tab:1} lists the electron candidate yields in ON data, the estimated
background contributions and the subsequent extracted signal for
the six overlapping momentum intervals.

  \begin{table}[hbpt!]
  \caption{The
    $B\rightarrow X_u e \nu_e$ endpoint and background yields in six
    momentum intervals, where the first error
    is statistical and the second is systematic. \label{tab:1}}
  \begin{center}
  \footnotesize
    \begin{tabular}{l|r|r|r}
      \hline
      $p_{\mathrm{CM}}\,(\gev/c)$
      & $2.4 - 2.6$
      & $2.3 - 2.6$
      & $2.2 - 2.6$
      \\
      \hline
      \hline
      $N_{\mathrm{ON}}$
      & $1741$
      & $3534$
      & $8854$
      \\\hline
      $\alpha N_{\mathrm{OFF}}$
      & $1166\pm 59 \pm  5$
      & $1878\pm 75 \pm  8$
      & $2743\pm 91 \pm 11$
      \\
      Fakes
      &  $12 \pm 2 \pm  4 $
      &  $43 \pm 4 \pm 14 $
      &  $85 \pm 5 \pm 28 $
      \\
      $N_{B\rightarrow J/\psi\rightarrow e}$
      & $40  \pm 3  \pm 4$
      & $94  \pm 5  \pm 8$ 
      & $191 \pm 7 \pm 17$
      \\
      $N_{B\rightarrow X \rightarrow e}$
      & $8 \pm  1 \pm 1$
      & $23 \pm 2 \pm 2$
      & $53 \pm 4 \pm 4$
      \\
      $N_{B\rightarrow X_c e \nu_e}$
      & $4   \pm 1  \pm 0$  
      & $345 \pm 11 \pm 23$ 
      & $3658 \pm 36 \pm 151$ 
      \\
      \hline
      $N_{B\rightarrow X_u e \nu_e}$
      &  $512  \pm  73 \pm   7$
      &  $1152 \pm  97 \pm  29$
      &  $2124 \pm 136 \pm 155$
      \\
    \hline
    \hline
    \multicolumn{4}{c}{} \\   
    \hline
    $p_{\mathrm{CM}}\,(\gev/c)$
    & $2.1 - 2.6$
    & $2.0 - 2.6$
    & $1.9 - 2.6$
    \\
    \hline
    \hline
    $N_{\mathrm{ON}}$
    & $23617$ 
    & $54566$
    & $104472$ 
    \\\hline
    $\alpha N_{\mathrm{OFF}}$
    & $3738\pm 106 \pm 15$
    & $4900\pm 121 \pm 20$
    & $6234\pm 137 \pm 25$  
    \\
    Fakes
    &  $93  \pm 6 \pm 34 $
    &  $131 \pm 7 \pm 40 $
    &  $143 \pm 9 \pm 47 $
    \\
    $N_{B\rightarrow J/\psi\rightarrow e}$
    & $336 \pm 9  \pm 29$
    & $562 \pm 12 \pm 49$
    & $880 \pm 15 \pm 77$ 
    \\
    $N_{B\rightarrow X \rightarrow e}$
    & $127 \pm 5  \pm 10$
    & $263 \pm 8  \pm 22$
    & $553 \pm 11 \pm 48$ 
    \\
    $N_{B\rightarrow X_c e \nu_e}$
    & $15494 \pm 73  \pm  437$
    & $42769 \pm 123 \pm  970$
    & $87705 \pm 180 \pm 1550$ 
    \\
    \hline
    $N_{B\rightarrow X_u e \nu_e}$
    &  $3830 \pm 201 \pm 439$
    &  $5941 \pm 291 \pm 972$
    &  $8957 \pm 395 \pm 1553$ 
    \\
    \hline
    \hline
  \end{tabular}
  \end{center}
\end{table}

\section{Extraction of the partial branching fraction}
The inclusive partial branching
fraction is found using
\begin{equation}
  \Delta \mathcal{B} = \frac{N(B\rightarrow X_u e
\nu)}{2N_{B\overline{B}}\epsilon_{\mathrm{MC}}}
\end{equation} 
where
$N_{B\overline{B}}=(29.4 \pm 0.4)\times 10^6$ 
and the overall selection efficiency is
$\epsilon_{\mathrm{MC}}$.
The systematic uncertainty on the efficiency includes effects
from tracking, electron
identification, event selection, or model criteria:
\begin{itemize}
\item
  The uncertainty on the track finding for our electron candidates
  is studied using the MC simulated track embedding method.
  Care is taken to consider all known sources of
  uncertainty in the MC simulation: magnetic field effects; CDC wire hit
  inefficiency; uncertainties in the material budget of the SVD
  and CDC; and drift time resolution effects in the CDC.
  The ratio of data to MC
  single track reconstruction efficiency is consistent with unity at
  the 1\% level. Accordingly, this uncertainty is assigned as the
  systematic error on the efficiency due to tracking;
\item
  The uncertainty in electron identification (ID) efficiency is
  measured using inclusive  $J/\psi$ events (The method implemented is
  similar to that
  described in Ref.~\cite{Hanagaki:2001fz}).
  The study involves reconstructing $J/\psi\rightarrow e^+e^-$ decays 
  with both tracks satisfying the same track requirements as those of the
  electron candidates considered for this analysis.
  We find excellent agreement of the MC simulation with
  data at the level of $2\%$ and subsequently assign this 
  as the systematic uncertainty on electron identification;
\item
  The effect of the main event selection criteria, namely those of
  $\mathcal{K}$ dependent \flow\, and  $\cos\theta_{thr}$ cuts,
  is assessed in two control samples. We fully reconstruct
  $B^+ \rightarrow \overline{D}^0 (\rightarrow K^+\pi^-) \rho^+
  (\rightarrow \pi^+\pi^0)$ decays. Here the kaon,
  disregarding particle identification, is assigned as the electron
  candidate, whilst the pion is regarded as the neutrino. In comparison
  to $B \rightarrow X_u e \nu_e$, the mass of the $D$ meson fixes $q^2=m_D^2$. 
  The data to MC ratio of the selection efficiency 
  is calculated as a function of CM momentum in the
  range $1.5-2.6$ GeV/$c$, in bins of 
  $0.05$ GeV/$c$; the best fit is achieved with a constant, which
  is found to be consistent with unity within 2\% uncertainty. We also fully reconstruct
  $J/\psi\rightarrow e^+e^-$ decays
  and subtract off backgrounds to
  yield $B\rightarrow J/\psi X$ decays. We assign the highest
  momentum electron from the $J/\psi$ decay to be the electron
  candidate. The remaining electron is regarded as the neutrino. The
  mass of the $J/\psi$-meson fixes $q^2=m^2_{J/\psi}$.
  The selection efficiency in this sample was measured as described
  above, and the best fit, which was also achieved with a constant, found
  data and MC agree to within 3\% uncertainty. Accordingly an overall
  uncertainty of 4\% is assigned as the
  systematic error due to event selection;
\item
  Model dependence is assessed using the four inclusive samples described
  above. The maximum shift in selection efficiency is assigned
  as the systematic uncertainty due to model dependence, and is
  dependent upon the particular HI region. It varies from 1.7\% to 3.4\%
  as the lower momentum limit is increased. 
\end{itemize}

The efficiencies for selecting electrons from $B \rightarrow X_u e \nu_e$ decays after
all selection criteria have been
applied are given in Table~\ref{tab:2}.
Our total efficiency decreases as the lower limit of the electron
momentum interval increases, an effect due mostly to the momentum
dependence of the $\mathcal{K}$ dependent \flow\, cut.

\begin{figure}
\includegraphics[width=1.00\textwidth]{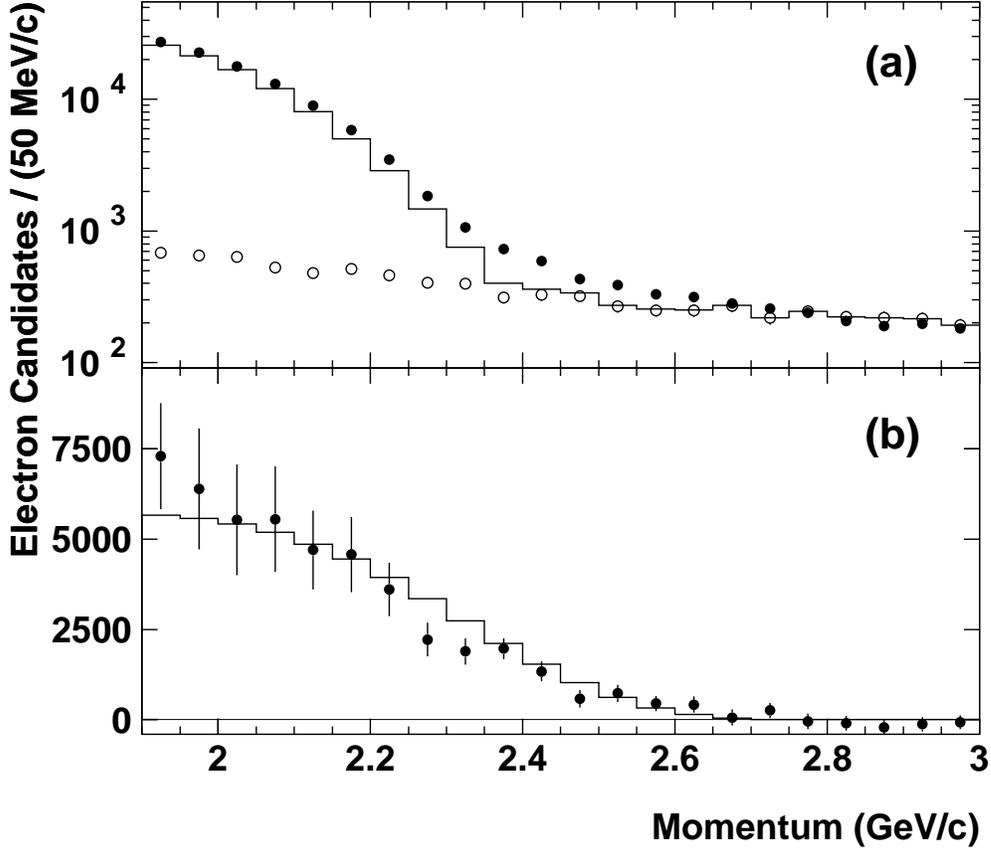}
\caption{The electron momentum spectrum in the $\Upsilon(4S)$ rest frame:
  (a) ON data (filled circles), scaled OFF data
  (open circles), sum of scaled OFF  data and estimated $B\overline B$ 
   backgrounds (histogram). (b) ON
  data after subtraction of
  backgrounds and correction for efficiency (filled circles) and 
model spectrum of
  $B \rightarrow X_u e \nu_e$ decays with final state radiation 
  (histogram, normalised to the data yield in the
  $1.9-2.6\,\mathrm{GeV}/c$ momentum range).}
\label{fig:1}       
\end{figure}

Figure~\ref{fig:1}(a) shows the ON and scaled OFF momentum spectra
along with the total background. 
Figure~\ref{fig:1}(b) shows the
ON spectrum after background subtraction and efficiency correction, 
revealing the contribution of $B\rightarrow
X_u e \nu_e$. The shape prescribed by the
inclusive model described earlier, with final state radiation, is also
shown. The partial branching fractions for each momentum interval
are given in Table~\ref{tab:2}; as the lower momentum limit is
decreased the uncertainty comes to be dominated, as expected, by the uncertainty in
the $B\rightarrow X_c e \nu_e$ background subtraction. Our partial
branching fraction measurements are consistent with those of CLEO and
have overall reduced uncertainties~\cite{Bornheim:2002du}.

\begin{table} 
  \caption{Branching fractions and extraction of $|V_{ub}|$(DFN method). The reconstruction efficiency, $\epsilon_{\mathrm{MC}}$, as
    calculated from Monte Carlo.
    The partial branching fractions, $\Delta\mathcal{B}_u(p)$,
    where the errors are from statistics and experimental systematics,
    respectively. The lepton momentum spectral fractions, $f_u$, where
    the first error is the combined statistical and systematic
    uncertainty, and the second error is the theoretical uncertainty in
    extracting shape function parameters from $B\rightarrow X_s\gamma$
    decays and applying this knowledge to $B\rightarrow X_u l \nu_l$
    decays. The correction due to the final state radiation loss
    is denoted $\delta_{\mathrm{RAD}}$.
    The full branching fractions, $\mathcal{B}(B\rightarrow X_u
    l \nu_l)$, where the first error is due to experimental uncertainty and
    the second is from $f_u$. The extracted values of $|V_{ub}|$: the first
    error is experimental; the second error is from $f_u$, combined
    statistical and systematic; the third error is from $f_u$ theory;
    and the last is from the application of the $|V_{ub}|$ formula given
    in Eqn~\ref{eq:vub}.   
}
\label{tab:2}
  \begin{center}
  \footnotesize
    \begin{tabular}{cccc}
    \hline
    $p_{\mathrm{CM}}\,(\gev/c)$
    & $\epsilon_{\mathrm{MC}}(\%)$ 
    &  $\Delta\mathcal{B}_u(p)(10^{-4})$
    & $f_u$\\
    \hline
    \hline
    $1.9-2.6$
    & $18.0 \pm 0.9$ 
    & $8.47 \pm 0.37 \pm 1.53$
    & $0.321 \pm 0.022 \pm 0.041$\\
    $2.0-2.6$
    & $17.6 \pm 0.9$ 
    & $5.74 \pm 0.28 \pm 0.98$
    & $0.246 \pm 0.020 \pm 0.042$\\
    $2.1-2.6$
    & $17.2 \pm 0.9$ 
    & $3.78 \pm 0.20 \pm 0.48$
    & $0.173 \pm 0.017 \pm 0.040$\\
    $2.2-2.6$
    & $16.6 \pm 0.9$ 
    & $2.17 \pm 0.14 \pm 0.20$
    & $0.109 \pm 0.013 \pm 0.034$\\
    $2.3-2.6$
    & $16.5 \pm 0.9$ 
    & $1.18 \pm 0.10 \pm 0.07$
    & $0.058 \pm 0.010 \pm 0.025$\\
    $2.4-2.6$
    & $16.2 \pm 1.0$ 
    & $0.53 \pm 0.07 \pm 0.03$
    & $0.025 \pm 0.006 \pm 0.014$\\
    \hline
    \hline
    $p_{\mathrm{CM}}\,(\gev/c)$
    & $\delta_{\mathrm{RAD}}$
    & $\mathcal{B}(B\rightarrow X_u e \nu_e)(10^{-3})$
    & $|V_{ub}|(10^{-3})$(DFN) \\
    \hline
    \hline
    $1.9-2.6$
    & $0.06 \pm 0.02$
    & $2.80 \pm 0.52 \pm 0.41$
    & $5.01 \pm 0.47 \pm 0.17 \pm 0.32 \pm 0.24$ \\
    $2.0-2.6$
    & $0.07 \pm 0.02$
    & $2.49 \pm 0.45 \pm 0.47$
    & $4.73 \pm 0.42 \pm 0.19 \pm 0.40 \pm 0.23$ \\
    $2.1-2.6$
    & $0.07 \pm 0.02$
    & $2.34 \pm 0.33 \pm 0.59$
    & $4.59 \pm 0.32 \pm 0.22 \pm 0.53 \pm 0.22$ \\
    $2.2-2.6$
    & $0.09 \pm 0.03$
    & $2.16 \pm 0.25 \pm 0.73$
    & $4.41 \pm 0.25 \pm 0.27 \pm 0.69 \pm 0.21$ \\
    $2.3-2.6$
    & $0.10 \pm 0.03$
    & $2.22 \pm 0.24 \pm 1.02$
    & $4.47 \pm 0.24 \pm 0.36 \pm 0.96 \pm 0.22$ \\
    $2.4-2.6$
    & $0.11 \pm 0.04$
    & $2.39 \pm 0.38 \pm 1.46$
    & $4.63 \pm 0.37 \pm 0.53 \pm 1.32\pm 0.22$ \\
    \hline
    \hline
 \end{tabular}
  \end{center}
\end{table}

\section{Extraction of $|V_{ub}|$ \label{sec:vubcleo}}
\subsection{DFN method}
The value of $|V_{ub}|$ is extracted using the formula given in Ref.~\cite{Group:2004cx}:
\begin{eqnarray}
\hspace{-5mm}  
\label{eq:vub}
  |V_{ub}| & = & 0.00424  \sqrt{
\frac
{\mathcal{B}(B\rightarrow X_u e \nu_e) }{0.002}
\frac
{1.604~\mathrm{ps}}
{\tau_B}
} \times (1.0 \pm 0.028_{\lambda_{1,2}} \pm 0.039_{m_b}),
\end{eqnarray}
which is an updated version of the expression given in
Ref.~\cite{Eidelman:2004wy}, and includes the latest measurements of
the heavy-quark expansion parameters~\cite{Aubert:2004aw}. 
We average the current world average neutral and charged
$B$-meson lifetimes to obtain $\tau_B=1.604\pm 0.011~\mathrm{ps}$~\cite{Eidelman:2004wy}.
To obtain the full inclusive rates for charmless semileptonic $B$-meson decay 
from our partial branching fractions, we must determine the spectral
fractions $f_u$ and the spectral distortion due to final state
radiation loss $\delta_{\mathrm{RAD}}$ such that
\begin{equation}
\mathcal{B}(B\rightarrow X_u e \nu_e) = \frac{\Delta\mathcal{B}(B\rightarrow X_u e
  \nu_e)}{f_u} (1+\delta_{\mathrm{RAD}}).
\end{equation}

The value of $\delta_{\mathrm{RAD}}$ is calculated from a comparison
of MC signal events generated with and without the PHOTOS algorithm
implemented. It has been the convention to assign a 10\% systematic
uncertainty on the correction
based on studies that compare the PHOTOS
performance with next-to-leading order
calculations in $B\rightarrow D e \nu_e$
decays~\cite{Richter-Was:1993ta}, since the study has yet to be extended to $B\rightarrow X_u e
\nu_e$ decays, we assign a larger uncertainty, equivalent to a third
of the size of the effect. The correction
factors for each HI region are given in Table~\ref{tab:2}.

The values of $f_u$ for the different momentum intervals are
determined with the DeFazio-Neubert prescription, using three
different forms of the shape function with
the parameters, $\Lambda^\mathrm{SF}$ and
$\lambda^\mathrm{SF}_1$, determined 
from fits to the Belle measured photon energy spectrum in
$B\rightarrow X_s \gamma$ decays~\cite{Koppenburg:2004fz,Limosani:2004jk}.
The resultant values of $f_u$ are given in Table~\ref{tab:2}, they
range from $3-32\%$ as the lower momentum limit is decreased.
The statistical uncertainty, averaged over each shape function form,
is determined from the half-difference of maximum and minimum $f_u$
found on the $\Delta\chi^2=1$ contour in
$(\Lambda^{\mathrm{SF}},\lambda^\mathrm{SF}_1)$ parameter space.
The systematic error stems from variation of the
scale used to evaluate the strong coupling $\alpha_s(\mu)$
($\mu=m_b/2,2m_b$)
and differences among
shape function forms. The theoretical uncertainty is obtained by varying
the parameters
by $\pm 20\%$, reflecting the fact that the
procedure is correct only to leading order in the HQET expansion that describes
the non-perturbative dynamics of $B$-mesons. Our variation
is twice that considered by CLEO ($\pm 10\%$). At the time of
their evaluation little was known about sub-leading and weak
annihilation 
effects, they have since been
predicted to be
large~\cite{Leibovich:2002ys,Bauer:2002yu,Neubert:2002yx} and are
better represented by a $\pm 20\%$ variation.
The resulting full branching fractions and extracted values
of $|V_{ub}|$ are given in Table~\ref{tab:2}. All the 
uncertainties contributing to $|V_{ub}|$ are summarised in
Table~\ref{tab:3} for each momentum interval.
As expected, as the lower momentum cutoff is decreased, the uncertainty
from $f_u$ that is due to theory, decreases, while the main 
experimental systematic, the estimation of $B\rightarrow X_c e
\nu_e$, increases, in line with its background contribution.
The best
overall precision ($13\%$) on $|V_{ub}|$, based on a sum in quadrature
of experimental and theoretical uncertainties, is found for the
$1.9-2.6~\mathrm{GeV}/c$ momentum
interval with
\begin{equation}
  |V_{ub}|=(5.01 \pm 0.47 \pm 0.17 \pm 0.32 \pm 0.24)\times 10^{-3},
\end{equation}
where the first error is from experiment, the second and third are due
to experiment and theory errors on $f_u$, respectively, and the last
is the uncertainty in applying the $|V_{ub}|$ extraction formula.

\begin{table}[htbp]
  \caption{Uncertainties contributing to the determination of
    $|V_{ub}|(10^{-3})$ (DFN method). The total error is obtained
    from a sum in quadrature.}
  \footnotesize
  \begin{center}
    \begin{tabular}{l|c|c|c|c|c|c} \hline 
       & \multicolumn{6}{c}{Momentum Interval
      ($\gev/c$)} \\
      Source of Uncertainty  & $1.9-2.6$  &  $2.0-2.6$    &  $2.1-2.6$
      &  $2.2-2.6$  &  $2.3-2.6$  &  $2.4-2.6$ \\ \hline \hline
      Statistical      & 0.11 & 0.12 &  0.12 &  0.14 &  0.19 &  0.33 \\
      $B\rightarrow X_c e \nu_e$ background        &0.43 &      0.39 &  0.26 &  0.16 &  0.04 &  0.00 \\
      Other $B$ background       & 0.03 &       0.03 &  0.03 &  0.04 &  0.03 &  0.03 \\
      Efficiency-detector          &0.12 &      0.11 &  0.11 &  0.10 &  0.11 &  0.12 \\
      Efficiency-model          &0.04 & 0.04 &  0.05 &  0.05 &  0.05 &  0.08 \\
      $N_{B\overline{B}}$              & 0.03 & 0.03 &  0.03 &  0.03 &  0.03 &  0.03 \\
      $\delta_{\mathrm{RAD}}$            &0.05 &        0.05 &  0.05 &  0.06 &  0.07 &  0.08 \\
      $f_u$ statistical          & 0.17 &       0.19 &  0.22 &  0.26 &  0.35 &  0.51 \\
      $f_u$ systematic            & 0.04 &      0.05 &  0.05 &  0.06 &  0.09 &  0.13 \\
      $f_u$ theory        &  0.32 &       0.40 &  0.53 &  0.69 &  0.96 &  1.32 \\
      $|V_{ub}|$ : $m_b^{\mathrm{kin}}(1\,\mathrm{GeV}),\,\lambda_{1,2}$        & 0.24 &       0.23 &  0.22 &  0.21 &  0.22 &  0.22 \\\hline
      Total          & 0.64 & 0.66 &  0.69 &  0.81 &  1.07 &  1.48 \\
      \hline \hline
    \end{tabular}
  \end{center}
  \label{tab:3} 
\end{table}

\subsection{BLNP method}
In this prescription $|V_{ub}|$ is obtained directly from the partial
branching                  
fraction $\Delta \mathcal{B}$, using the formula  
\begin{equation}
  |V_{ub}| = \sqrt{\frac{\Delta
   \mathcal{B}(1+\delta_{\mathrm{RAD}})}{\tau_B}\frac{1}{R}},
  \label{eqn:5}
\end{equation}	                                                     
where $\tau_B$ and $\delta_{\mathrm{RAD}}$ are as the same as
described previously, and $R$ is                         
the theoretical prediction of the partial rate of $B\rightarrow
X_u l \nu$ decay in units of $|V_{ub}|^2\mathrm{ps}^{-1}$                        
for a given momentum region.
The implementation of the BLNP method relies on a model for the
leading order shape function that is constrained by HQET parameters;
the mass and average momentum squared of the $b$-quark,
$m_b(\mathrm{SF})$ and $\mu_\pi^2(\mathrm{SF})$, respectively,
as defined in the shape function scheme (SF). They are set to
$m_b(\mathrm{SF})=(4.52 \pm 0.07)\,\mathrm{GeV}/c^2$ and
$\mu_\pi^2(\mathrm{SF})=(0.27 \pm 0.13)\,\mathrm{GeV}^2/c^2$,
as derived from the measurement procedure 
that is described in the introduction. 

Table~\ref{tab:4} gives the $R$ and $|V_{ub}|$ values for
the overlapping momentum intervals.
The first error on $R$ is the experimental uncertainty on the leading
order shape function, which is our own estimation calculated as the
half-difference of minimum
and maximum $R$ values obtained from a set of shape function
parameters which lie on the $\Delta\chi^2=1$ contour. The second error on $R$
is a theoretical uncertainty
stemming from the variation of the matching scales $\mu_i,\bar{\mu},\mu_h$,
sub-leading shape function models and the weak annihilation effect,
where the latter effect is constant ($\pm 1.40$) for all momentum
intervals~\cite{Lange:2005yw}.

Our most precise value, which has an overall
uncertainty of $13\%$ as based on the sum in
quadrature of all the uncertainties,
\begin{equation}
   |V_{ub}|=(5.08 \pm 0.47 \pm 0.42^{\,+\,0.26}_{\,-\,0.23})\times 10^{-3},
\end{equation}
is found for the $1.9-2.6~\mathrm{GeV}/c$ momentum interval. When the
shape function parameters and consequently $R$ are better determined,
$|V_{ub}|$ can be re-calculated from the partial branching fraction
measurements presented in Table~\ref{tab:2}.  

\begin{table} 
  \caption{Predicted partial rate $R$ for $B\rightarrow X_u l
    \nu_l$ and extracted value of
    $|V_{ub}|$(BLNP method).
    The first error in $R$ is the shape function
    error stemming from the uncertainty in the knowledge
    of HQET parameters and the second is a theoretical uncertainty  
    stemming from the variation of the matching scales $\mu_i,\bar{\mu},\mu_h$,
    subleading shape function models and the weak annihilation effect.
    The first error in
    $|V_{ub}|$ is the experimental error, and the remaining errors are
    those propagated from $R$, respectively.}
  \label{tab:4}
  \begin{center}
    \begin{tabular}{c|r|c}
      \hline
      $p_{\mathrm{CM}}\,(\gev/c)$
      & \multicolumn{1}{c|}{$R\,(|V_{ub}|^2\mathrm{ps}^{-1})$} 
      & $|V_{ub}|\,(10^{-3})$(BLNP)\\
      \hline\hline
      $1.9-2.6$
      & $21.69\pm 3.62^{\,+\,2.18}_{\,-\,1.98}$  
      & $5.08 \pm 0.47 \pm 0.42^{\,+\,0.26}_{\,-\,0.23}$  \\ 
      $2.0-2.6$ 
      & $16.05 \pm 3.05^{\,+\,1.83}_{\,-\,1.72}$  
      & $4.87 \pm 0.43 \pm 0.46^{\,+\,0.28}_{\,-\,0.26}$  \\ 
      $2.1-2.6$
      & $10.86 \pm 2.51^{\,+\,1.61}_{\,-\,1.57}$  
      & $4.83 \pm 0.33 \pm 0.56^{\,+\,0.36}_{\,-\,0.35}$  \\ 
      $2.2-2.6$
      & $6.46 \pm 1.54^{\,+\,1.54}_{\,-\,1.53}$  
      & $4.77 \pm 0.26 \pm 0.57^{\,+\,0.57}_{\,-\,0.56}$  \\ 
      $2.3-2.6$
      & $3.15 \pm 0.88^{\,+\,1.55}_{\,-\,1.54}$  
      & $5.07 \pm 0.71 \pm 0.52^{\,+\,1.25}_{\,-\,1.24}$  \\ 
      $2.4-2.6$
      & $1.12 \pm 0.39^{\,+\,1.48}_{\,-\,1.48}$  
      & $5.70 \pm 1.00 \pm 0.67^{\,+\,3.77}_{\,-\,3.76}$  \\ 
      \hline
      \hline
    \end{tabular}
  \end{center}

\end{table}

\section{Summary}
We have measured the inclusive charmless
semileptonic $B$-meson decay branching ratio in six overlapping momentum intervals that encompass
the endpoint of the electron momentum spectrum. These 
included a momentum interval with a minimum lower momentum cutoff of
$1.9\,\mathrm{GeV}/c$, from which  
the partial
branching fraction was measured to be $\Delta \mathcal{B}=(8.47
\pm 0.37 \pm 1.53)\times 10^{-4}$.
We have extracted $|V_{ub}|$ using both the DFN and BLNP methods, but           
we adopt the results of the latter method since it is more advanced.
The most precise $|V_{ub}|$ value was extracted from the decay rate in the
$1.9-2.6~\mathrm{GeV}/c$ momentum interval and found to be
$|V_{ub}|=(5.08 \pm 0.47 \pm 0.42^{\,+\,0.26}_{\,-\,0.23})\times 10^{-3}$.
Owing to updated knowledge of background shapes and normalisations, as
well as the improvement in the theoretical prediction of the decay
rates for $B\rightarrow X_u e\nu_e$ and $B\rightarrow X_s \gamma$
decays, the precision of the present measurement is better than
that of the previous endpoint measurement by CLEO~\cite{Bornheim:2002du}.
Although endpoint methods have not been preferred for a precison
determination of
$|V_{ub}|$ from inclusive decays~\cite{Bauer:2001rc,Bosch:2004bt},
the results presented in this letter for the momentum interval $1.9-2.6\,\mathrm{GeV}/c$
are competitive in precision with measurements that have utilised the favoured kinematic regions of hadronic mass and
dilepton mass squared~\cite{Aubert:2003zw,Kakuno:2003fk}.
This competitiveness is due to a minimum lower momentum cutoff of $1.9\,\mathrm{GeV}/c$.
Our results also, independent of the extracted value of $|V_{ub}|$,
help to bound theoretical uncertainties that in general are
encountered in all $|V_{ub}|$ extractions from inclusive charmless
semileptonic $B$-meson decays, for example, those relating to
quark-hadron duality and the weak annihilation
effect~\cite{Gibbons:2004dg}.

The comparison of our result with other experimental measurements
of $|V_{ub}|$~\cite{Bornheim:2002du,Aubert:2003zw,Kakuno:2003fk}
must be made on a consistent basis, that is, the extraction of
$|V_{ub}|$ from a partial branching fraction measurement needs to be
performed using a common theoretical framework with common inputs.

\section*{ACKNOWLEDGEMENTS}
We are grateful to Matthias Neubert, Bjorn Lange, Gil Paz and Stefan
Bosch for very helpful discussions, correspondences, explanations and
for providing us with their theoretical computations implemented in an
inclusive generator.
We thank the KEKB group for the excellent operation of the
accelerator, the KEK cryogenics group for the efficient
operation of the solenoid, and the KEK computer group and
the National Institute of Informatics for valuable computing
and Super-SINET network support. We acknowledge support from
the Ministry of Education, Culture, Sports, Science, and
Technology of Japan and the Japan Society for the Promotion
of Science; the Australian Research Council, the
Australian Department of Education, Science and Training, Australian
Postgraduate Award and the
David Hay Postgraduate Writing-Up Award;
the National Science Foundation of China under contract
No.~10175071; the Department of Science and Technology of
India; the BK21 program of the Ministry of Education of
Korea and the CHEP SRC program of the Korea Science and
Engineering Foundation; the Polish State Committee for
Scientific Research under contract No.~2P03B 01324; the
Ministry of Science and Technology of the Russian
Federation; the Ministry of Higher Education, Science and Technology of
the Republic of Slovenia;  the Swiss National Science Foundation; the National Science Council and
the Ministry of Education of Taiwan; and the U.S.\
Department of Energy. 

\bibliographystyle{elsart-num}

\begin{thebibliography}{00}

\bibitem{Abe:2001xe}
K.~Abe {\it et al.}  [Belle Collaboration],
Phys.\ Rev.\ Lett.\  {\bf 87} (2001) 091802.

K.~Abe {\it et al.}  [Belle Collaboration],
Phys.\ Rev.\ D {\bf 66} (2002) 032007.

K.~Abe {\it et al.}  [Belle Collaboration],
Phys.\ Rev.\ D {\bf 66} (2002) 071102.

\bibitem{Aubert:2001nu}
B.~Aubert {\it et al.}  [BABAR Collaboration],
Phys.\ Rev.\ Lett.\  {\bf 87} (2001) 091801.

B.~Aubert {\it et al.}  [BABAR Collaboration],
Phys.\ Rev.\ D {\bf 66} (2002) 032003.

B.~Aubert {\it et al.}  [BABAR Collaboration],
Phys.\ Rev.\ Lett.\  {\bf 89} (2002) 201802.
\bibitem{KM}
  M.~Kobayashi and T.~Maskawa,
  Prog.\ Theor.\ Phys.\  {\bf 49} (1973) 652.


\bibitem{Fulton:1989pk}
  R.~Fulton {\it et al.}  [CLEO Collaboration],
  Phys.\ Rev.\ Lett.\  {\bf 64} (1990) 16.
  
\bibitem{Bartelt:1993xh}
J.~Bartelt {\it et al.}  [CLEO Collaboration],
Phys.\ Rev.\ Lett.\  {\bf 71} (1993) 4111.

\bibitem{Bornheim:2002du}
  A.~Bornheim {\it et al.}  [CLEO Collaboration],
  Phys.\ Rev.\ Lett.\  {\bf 88} (2002) 231803.
  
\bibitem{Albrecht:1989qv}
H.~Albrecht {\it et al.}  [ARGUS Collaboration],
Phys.\ Lett.\ B {\bf 234} (1990) 409.

\bibitem{Aubert:2003zw}
B.~Aubert {\it et al.}  [BABAR Collaboration],
Phys.\ Rev.\ Lett.\  {\bf 92} (2004) 071802.

\bibitem{Kakuno:2003fk}
H.~Kakuno {\it et al.}  [BELLE Collaboration],
Phys.\ Rev.\ Lett.\  {\bf 92} (2004) 101801.

\bibitem{Acciarri:1998if}
M.~Acciarri {\it et al.}  [L3 Collaboration],
Phys.\ Lett.\ B {\bf 436} (1998) 174.

\bibitem{Barate:1998vv}
R.~Barate {\it et al.}  [ALEPH Collaboration],
Eur.\ Phys.\ J.\ C {\bf 6} (1999) 555.

\bibitem{Abreu:2000mx}
P.~Abreu {\it et al.}  [DELPHI Collaboration],
Phys.\ Lett.\ B {\bf 478} (2000) 14.

\bibitem{Abbiendi:2001qx}
G.~Abbiendi {\it et al.}  [OPAL Collaboration],
Eur.\ Phys.\ J.\ C {\bf 21} (2001) 399.


\bibitem{Group:2004cx}
H.~F.~A.~Group,
arXiv:hep-ex/0412073.


\bibitem{Bosch:2004th}
S.~W.~Bosch, B.~O.~Lange, M.~Neubert and G.~Paz,
Nucl.\ Phys.\ B {\bf 699} (2004) 335.

\bibitem{Neubert:2004dd}
  M.~Neubert,
  Eur.\ Phys.\ J.\ C {\bf 40} (2005) 165.
  
\bibitem{Bosch:2004cb}
S.~W.~Bosch, M.~Neubert and G.~Paz,
JHEP {\bf 0411} (2004) 073.

\bibitem{Neubert:2004cu}
M.~Neubert,
arXiv:hep-ph/0411027.

\bibitem{Neubert:2004sp}
  M.~Neubert,
  Phys.\ Lett.\ B {\bf 612} (2005) 13.

\bibitem{DeFazio:1999sv}
  F.~De Fazio and M.~Neubert,
  JHEP {\bf 9906} (1999) 017.
   

\bibitem{Koppenburg:2004fz}
P.~Koppenburg {\it et al.}  [Belle Collaboration],
Phys.\ Rev.\ Lett.\  {\bf 93} (2004) 061803.

\bibitem{Bigi:1993ex}
  I.~I.~Y.~Bigi, M.~A.~Shifman, N.~G.~Uraltsev and A.~I.~Vainshtein,
    Int.\ J.\ Mod.\ Phys.\ A {\bf 9} (1994) 2467.

\bibitem{Neubert:1993um}
M.~Neubert,
  Phys.\ Rev.\ D {\bf 49} (1994) 4623.

\bibitem{Limosani:2004jk}
A.~Limosani and T.~Nozaki  [Heavy Flavor Averaging Group],
arXiv:hep-ex/0407052.

\bibitem{Lange:2005yw}
  B.~O.~Lange, M.~Neubert and G.~Paz,
 arXiv:hep-ph/0504071 and private communication with B.~O.~Lange, M.~Neubert and G.~Paz.    
        

\bibitem{KEKB}
  S.~Kurokawa,
  Nucl.\ Inst rum.\ Meth.\ A {\bf 499} (2003) 1., and other papers included in this Volume.
  
\bibitem{bellenim}
A.~Abashian {\it et al.}  [Belle Collaboration], 
     Nucl.\ Instrum.\ Meth.\ A {\bf 479} (2002) 117.
       
       
\bibitem{Scora:1995ty}
  D.~Scora and N.~Isgur,
  Phys.\ Rev.\ D {\bf 52} (1995) 2783.
  
%
\bibitem{Fox:1978vu}
G.~C.~Fox and S.~Wolfram,
Phys.\ Rev.\ Lett.\  {\bf 41} (1978) 1581.

\bibitem{Hanagaki:2001fz}
  K.~Hanagaki, H.~Kakuno, H.~Ikeda, T.~Iijima and T.~Tsukamoto,
    Nucl.\ Instrum.\ Meth.\ A {\bf 485} (2002) 490.
  
\bibitem{BelleMC}
Events are generated with the CLEO QQ generator, see \\
http://www.lns.cornell.edu/public/CLEO/soft/QQ; \\
the detector response is simulated with GEANT, R.~Brun {\it et al.}, GEANT 3.21 CERN Report DD/EE/84-1, 1984.


\bibitem{Eidelman:2004wy}
  S.~Eidelman {\it et al.}  [Particle Data Group],
  Phys.\ Lett.\ B {\bf 592} (2004) 1.


\bibitem{hqet}
Form factors measured in
J.~E.~Duboscq {\it et al.}  [CLEO Collaboration],
  Phys.\ Rev.\ Lett.\  {\bf 76} (1996) 3898,
are used as input to the 
Heavy Quark Effective Theory, see, e.g., 
  M.~Neubert,
  Phys.\ Rept.\  {\bf 245} (1994) 259.
   
\bibitem{Barlow:dm}
R.~J.~Barlow and C.~Beeston,
Comput.\ Phys.\ Commun.\  {\bf 77} (1993) 219.

\bibitem{Barberio:1993qi}
  E.~Barberio and Z.~Was,
  Comput.\ Phys.\ Commun.\  {\bf 79} (1994) 291.


\bibitem{Goity:xn}
J.~L.~Goity and W.~Roberts,
Phys.\ Rev.\ D {\bf 51} (1995) 3459.


\bibitem{Aubert:2004aw}
B.~Aubert {\it et al.}  [BABAR Collaboration],
Phys.\ Rev.\ Lett.\  {\bf 93} (2004) 011803.

\bibitem{Richter-Was:1993ta}
E.~Richter-Was,
Z.\ Phys.\ C {\bf 61} (1994) 323.


\bibitem{Leibovich:2002ys}
  A.~K.~Leibovich, Z.~Ligeti and M.~B.~Wise,
  Phys.\ Lett.\ B {\bf 539} (2002) 242.


\bibitem{Bauer:2002yu}
C.~W.~Bauer, M.~Luke and T.~Mannel,
Phys.\ Lett.\ B {\bf 543} (2002) 261.

\bibitem{Neubert:2002yx}
M.~Neubert,
Phys.\ Lett.\ B {\bf 543} (2002) 269.

\bibitem{Bauer:2001rc}
  C.~W.~Bauer, Z.~Ligeti and M.~E.~Luke,
  Phys.\ Rev.\ D {\bf 64} (2001) 113004.

\bibitem{Bosch:2004bt}
  S.~W.~Bosch, B.~O.~Lange, M.~Neubert and G.~Paz,
  Phys.\ Rev.\ Lett.\  {\bf 93} (2004) 221801.

\bibitem{Gibbons:2004dg}
  L.~Gibbons  [CLEO Collaboration],
  AIP Conf.\ Proc.\  {\bf 722} (2004) 156.

\end{thebibliography}

\end{document}